\documentclass [aps,pre,twocolumn]{revtex4-1} 
\usepackage{lmodern}
\usepackage{graphicx}  
\usepackage{dcolumn}  
\usepackage{bm}       
\usepackage{amssymb}   
\usepackage{braket}
\usepackage{mathrsfs}
\usepackage{amsmath}
\usepackage{float}
\usepackage{color}
\usepackage[]{cancel}
\usepackage[dvipsnames]{xcolor}
\usepackage[hidelinks]{hyperref}
\hypersetup{colorlinks,linkcolor={NavyBlue}, citecolor={OliveGreen}, urlcolor={NavyBlue}}  
\usepackage{bookmark}

\begin{document}
\title{Confinement-induced Self-Pumping in 3D Active Fluids}
\author{Minu Varghese}
\affiliation{Martin Fisher School of Physics, Brandeis University, Waltham, MA 02453, USA}
\author{Arvind Baskaran}
\affiliation{Martin Fisher School of Physics, Brandeis University, Waltham, MA 02453, USA}
\author{Michael F. Hagan}
\affiliation{Martin Fisher School of Physics, Brandeis University, Waltham, MA 02453, USA}
\author{Aparna Baskaran}
\affiliation{Martin Fisher School of Physics, Brandeis University, Waltham, MA 02453, USA}
\begin{abstract}
Two dimensional active fluids display a transition from turbulent to coherent flow upon decreasing the size of the confining geometry. A recent experiment suggests that the behavior in three dimensions is remarkably different; emergent flows transition from turbulence to coherence upon \emph{increasing} the confinement height to match the width. Using a simple hydrodynamic model of a suspension of extensile rod-like units, we provide the theoretical explanation for this puzzling behavior. Furthermore, using extensive numerical simulations supported by theoretical arguments, we map out the conditions that lead to coherent flows and elucidate the critical role played by the aspect ratio of the confining channel. The mechanism that we identify applies to a large class of symmetries and propulsion mechanisms, leading to a unified set of design principles for self-pumping 3D active fluids. 
\end{abstract}
\maketitle
Active matter describes systems whose constituent particles consume energy to drive motion or generate forces \cite{Ramaswamy2010,Marchetti2013,Elgeti2015,Bechinger2016,Zottl2016,Elgeti2016,Yoshinaga2017,Saintillan2018,Seifert2019,Magistris2015,Doostmohammadi2018,Fodor2018,Baer2020,Shaebani2020}.
Being continuously driven far from equilibrium, active materials can exhibit spectacular spatiotemporal behaviors not possible in equilibrium systems. One such example is `self-pumping' flow, in which confining an active fluid leads to a spontaneous coherent flow, enabling advected material transport in the absence of any external driving such as a pressure gradient \cite{Voituriez2005}. Harnessing coherent spontaneous flow at scales relevant to practical devices would enable converting particle-scale chemical energy into macroscopic productive work, and thus has tremendous potential for practical applications. However, the mechanisms that drive and control self-pumping of confined active fluids are insufficiently understood to design such devices.

Spontaneous flow has been reported in diverse active fluids, including suspensions of microswimmers \cite{Woodhouse2012,Wioland2013,Wioland2016,Baskaran2009,Lushi2014,Koch2011,Lopez2015,Sokolov2012}, cell monolayers \cite{Duclos2018,Xi2019}, and active gels built from subcellular components \cite{Needleman2017,Julicher2007,Sanchez2012,Wu2017}. At the microscopic level, some of these systems have polar symmetry \cite{Woodhouse2012,Wioland2016,Baskaran2009,Lushi2014,Koch2011,Lopez2015,Sokolov2012}, and others have nematic symmetry \cite{Duclos2018,Xi2019,Needleman2017,Julicher2007,Sanchez2012,Wu2017}. Nevertheless, a unifying feature of all these systems is that they are composed of force dipoles in a fluid. In this work, we show that flow generated by force dipoles together with flow alignment and confinement give rise to system size dependent length scales in the structure of the flow, which can be harnessed to induce coherent flows in diverse systems. In particular, we uncover the underlying physics of aspect-ratio-dependent coherent flows observed in dilute isotropic suspensions of extensile microtubule bundles powered by ATP-driven molecular motors \cite{Wu2017}. 

Most previous theoretical studies of active matter, with a few notable exceptions\cite{Woodhouse2012,Thampi2015}, have considered models with polar self-propelled constituents\cite{Saintillan2007,Saintillan2008,Ishikawa2009,Subramanian2009,Baskaran2009,Hohenegger2010,Koch2011,Saintillan2012,Theillard2017,Saintillan2018,Mehandia2008,Hernandez-Ortiz2005}, or mutually aligning nematic constituents \cite{Ramaswamy2010,Marchetti2013,Elgeti2015,Bechinger2016,Zottl2016,Elgeti2016,Yoshinaga2017,Saintillan2018,Seifert2019,Needleman2017,Magistris2015,Doostmohammadi2018,Fodor2018,Baer2020,Shaebani2020}. Therefore, at least one of these ingredients --- self-propulsion or aligning interactions --- is often thought to be required for spontaneous flows. However, experiments by Wu et al. \cite{Wu2017} demonstrated emergent flows in meter long channels with no orientational order, i.e. well below the isotropic-nematic (IN) transition density of the microtubule bundles. Moreover, the coherence of the flow was non-monotonic as the size of the channel was varied, occurring only in channels with low aspect ratio cross-sections: $1/2\le H/W \le 2$, with $H$ and $W$ the height and width of the channel. This transition to spontaneous flow is thus an intrinsically 3D behavior, which cannot be explained by 2D models of active nematics \cite{Opathalage2019,Doostmohammadi2019,Shendruk2017,Hardouin2019}.

Here, we show that a minimal theoretical description of extensile microtubules below their IN transition, which includes only the force-dipole and flow-aligning character of the bundles, exhibits a transition to self-pumping flow that depends critically on dimensionality. We then reveal the physical origins of this behavior by formulating the aspect ratio dependence in terms of a confinement-induced length scale in the structure of the flow. Finally, we propose design principles for generating self-pumping flows in active fluids. The simplicity of our model implies that these design principles apply to a wide variety of seemingly disparate active matter systems.

\emph{Model.} Consider $N$ non-interacting ellipsoids, each of length $l$ and diameter $b$ suspended in a fluid in $d$ dimensions. The center of mass of ellipsoid $i$, $\vec{r}_i$, evolves as $\partial_t\vec{r}_i=\vec{u}(\vec{r}_i)+\sqrt{2\kappa}\ \vec{\eta}^T(\vec{r}_i)$, where $\vec{u}(\vec{r})$ is the fluid velocity at position $\vec{r}$, $\eta^T_\alpha$ is a stochastic Gaussian white noise, and $\kappa$ is the translational diffusion constant. Let the axis of orientation of this ellipsoid, $\pm \hat{\nu}_i$ be defined along its axis of symmetry (length). In low Reynolds number flows, the time evolution of $\hat{\nu}_i$ is given by \cite{Taylor1923,Bretherton1962}
\begin{align}
\partial _{t}\hat{\nu}_i=\boldsymbol{\Omega }(\vec{r}_i)\cdot \hat{\nu}_i+\lambda \left( \bm{I}-\hat{\nu}_i\hat{\nu}_i\right) \cdot
\mathbf{E}(\vec{r}_i)\cdot \hat{\nu}_i\nonumber\\
\hspace{20pt}+\sqrt{\gamma/2} \left( \bm{I}-\hat{\nu}_i\hat{\nu}_i\right)\cdot\ \vec{\eta}^R(\vec{r}_i)\label{microdyn}
\end{align}
where $\Omega_{\alpha\beta}=\frac{1}{2}(\partial_\beta u_\alpha-\partial_\alpha u_\beta)$ and $E_{\alpha\beta}=\frac{1}{2}(\partial_\beta u_\alpha+\partial_\alpha u_\beta)$ represent spatial variations in the flow, $\lambda=\frac{l^2-b^2}{l^2+b^2}$, $\vec{\eta}^R$ is a Gaussian white noise, and $\gamma$ is the rotational diffusion constant.

Suppose the ellipsoids are \emph{active}, pushing outward along their axes with a force $f$ and generating low Reynolds number flows around them. The emergent flow $\vec{u}$ is then a solution of the driven Stoke's equation,
$\eta \nabla^2\vec{u}(\vec{r})-\vec{\nabla}p=f\sum_{i}\hat{\nu}_i[\delta(\vec{r}-\vec{r}_i-l/2\ \hat{\nu}_i)-\delta(\vec{r}-\vec{r}_i+l/2\ \hat{\nu}_i)]$, \cite{Simha2002} where 
$\eta$ is the coefficient of viscosity and the mechanical pressure $p$ is such that flows are incompressible: $\vec{\nabla} \cdot \vec{u}=0$.

We can coarse grain this microscopic model to derive the dynamical description of this fluid on long length scales. This gives the dynamics of the coarse-grained nematic order, $\bm{Q}=\braket{\hat{\nu}\hat{\nu}-\frac{\bm{I}}{d}}$, to be of the form (see SI for derivation \cite{Doi1988,VanKampen2007,Baskaran2010})
 \begin{align}
&\hspace{-7pt}\partial_t\bm{Q}+\vec{u}\cdot\vec{\nabla}\bm{Q}+\bm{Q}\cdot\bm{\Omega}-\bm{\Omega}\cdot\bm{Q}=-\gamma \bm{Q} +\kappa \nabla^2\bm{Q}\nonumber\\
&\hspace{-10pt}+\lambda\left[\frac{2}{d}\bm{E}+\bm{Q}\cdot\bm{E}+\bm{E}\cdot\bm{Q}-\frac{2}{d}\ tr(\bm{Q}\cdot\bm{E})\bm{I}\right]\label{Qdynamics}
\end{align}
with $\eta \nabla^2\vec{u}-\vec{\nabla}p=\alpha\ \vec{\nabla} \cdot \bm{Q}$ and $\alpha=fl/2$. Note that our model is purely kinematic. Including passive nematic stresses in Eq. (\ref{Qdynamics}) would yield the Leslie-Erickson model for liquid crystals below the IN transition to linear order in $\bm{Q}$. This would decrease the effective activity \cite{Hatwalne2004}, but not qualitatively change the phenomenology discussed here.

\begin{figure}
\includegraphics[scale=0.15]{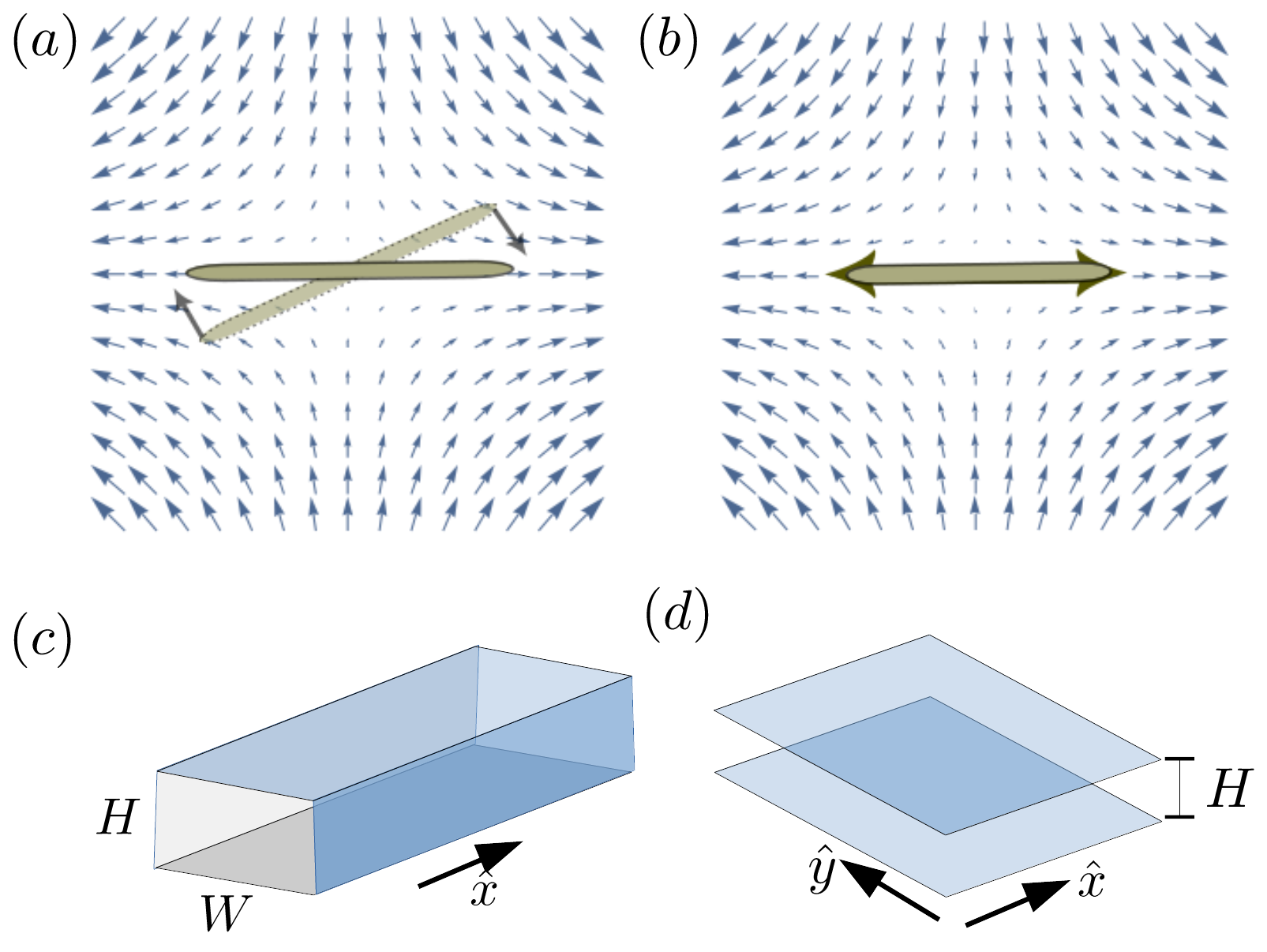}
 \caption{(a) A slender rod aligns with the extensional axis of a shear flow, (b) Flow field produced by an extensile rod, (c) confinement geometry for coherent flows, (d) confinement geometry for vortex size measurement}\label{schematic}
\end{figure}

\emph{Mechanism for spontaneous flow.} Since $\gamma>0$, the only homogeneous state admitted by the model is $\bm{Q}=0,\ \vec{u}=0$. Consider fluctuations about this stationary isotropic state, $Q_{\alpha\beta}(\vec{r})=\int d\vec{k}\ \tilde{Q}_{\alpha\beta}(\vec{k})e^{-i\vec{k}.\vec{r}}$, and $ u_{\alpha}(\vec{r})=\int d\vec{k}\ \tilde{u}_{\alpha}(\vec{k})e^{-i\vec{k}.\vec{r}}$. The effective dynamics of the flow is given by $ \partial_t\tilde{u}_\beta=\frac{i\alpha \hat{k}_\nu}{\eta k}(\delta_{\beta \mu}-\hat{k}_\beta \hat{k}_\mu)\partial_t\tilde{Q}_{\mu \nu}$. Up to linear order in perturbations, $
\partial_t\tilde{Q}_{\alpha\beta}=-(\gamma+\kappa k^2)\tilde{Q}_{\alpha\beta}-\frac{i\lambda}{d}(k_\alpha \tilde{u}_\beta+k_\beta \tilde{u}_\alpha)$. Eliminating $\tilde{Q}_{\alpha\beta}$, the effective linear dynamics of the flow is given by $
 \partial_t\tilde{u}_\beta=\left(-\gamma-\kappa k^2+\frac{\lambda\alpha}{d\eta}\right)\tilde{u}_\beta$. 
 Thus, the stationary isotropic state is destabilized by long wavelength perturbations if $\alpha\lambda>d\eta \gamma$, i.e., if the ellipsoids are rod-like ($\lambda>0$) and the forces exerted by them are sufficiently extensile ($\alpha>0$), or if the ellipsoids are discoidal ($\lambda<0$) and the forces exerted by them are sufficiently contractile ($\alpha<0$) \cite{Baskaran2009,Lushi2014,Koch2011,Saintillan2018,Thampi2015,Lopez2015,Chandragiri2019,Chandragiri2020}. 

 In this article, we will focus on extensile rod-like ellipsoids, to be consistent with experiments \cite{Wu2017}. The phenomenology of contractile discoidal objects is identical, so we will not consider them separately. The emergence of spontaneous flows in a suspension of extensile rod-like objects can be ascribed to the following microscopic mechanism: (1) rod-like objects orient along the extensional axis of shear flows (Fig. \ref{schematic}a), while (2) shear flows generated by extensile rods have an extensional axis that coincides with the the rod orientation (Fig. \ref{schematic}b). If the relevant active timescale, $\tau_{\rm a}=3\eta/\lambda\alpha$, is shorter than the timescale for loss of order due to rotational diffusion, $\tau_{\rm r}=1/\gamma$, flows and nematic order at hydrodynamic scales arise spontaneously \cite{Koch2011, Thampi2015}. Thus, a positive reinforcement between shear alignment and extensile active flows destroys the isotropic state and generates the spontaneous flows discussed hereafter.
 
In the rest of this article, we explore how confining walls can structure the spontaneous flows that are generated by the active shear alignment instability described above (see SI for full equations and numerical method \cite{Varghese2021,Zhao2016,Vanka1986}). The simplest theoretical set up that can sustain coherent flows is a 2D active fluid confined in a channel of width W. Above the flow alignment instability, strong confinements give rise to unidirectional coherent flows. As the width of the channel is increased, the flow becomes undulatory and finally turbulent. This is true regardless of anchoring at the walls (SI), and the observed steady states observed are similar to those discussed in the literature \cite{Voituriez2005,Edwards2009,Shendruk2017} for flows arising from the instability of orientationally ordered systems \cite{Simha2002, Voituriez2005}. However, this phenomenology does not trivially extend to 3D.

\begin{figure}
\centering
 \includegraphics[scale=0.14]{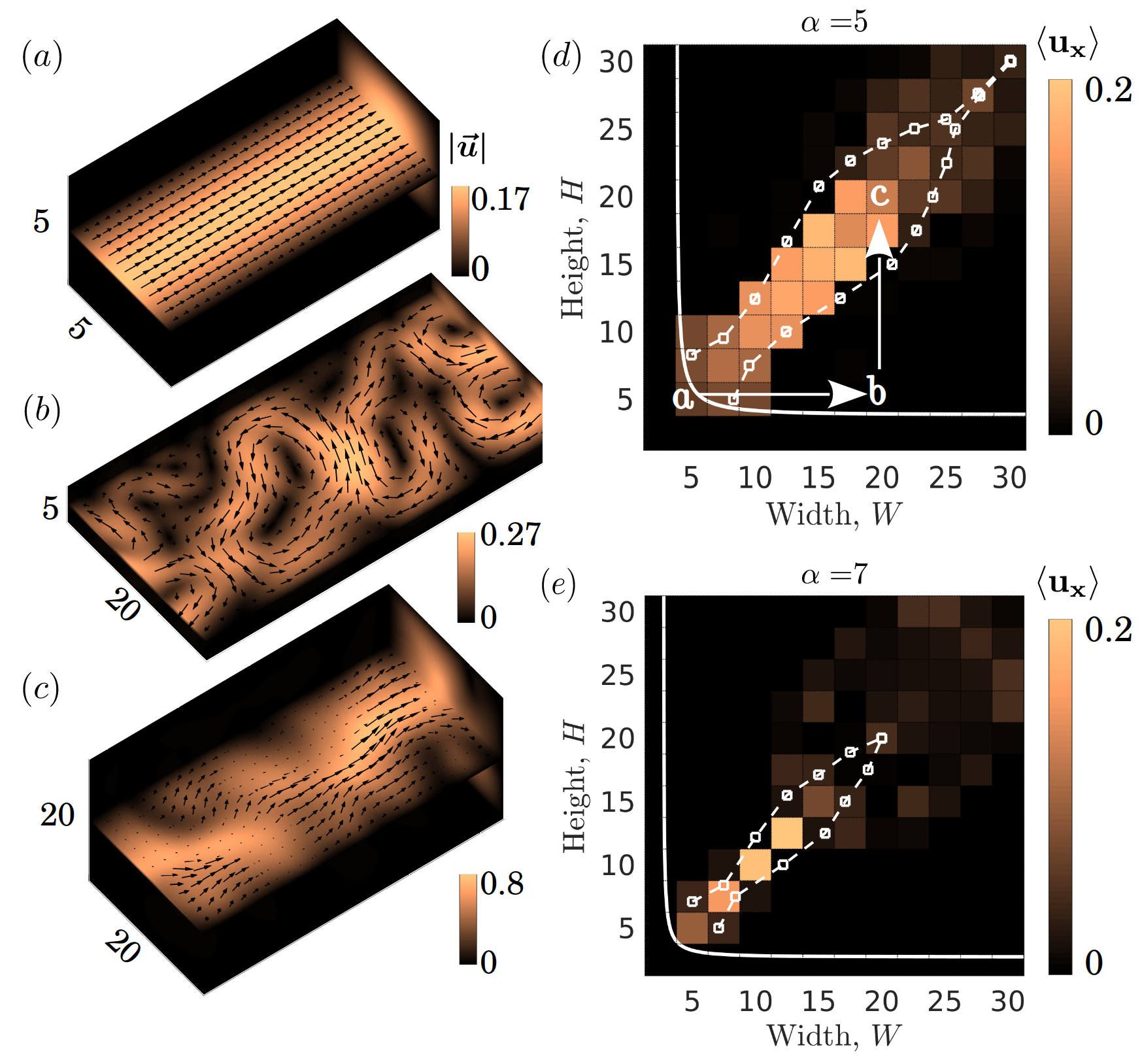}
 \caption{Net flow along the channel varies non-monotonically on monotonically increasing the cross-sectional area of the channel along the representative path (shown by arrows) in (d) \textbf{(a-c)} Instantaneous flow profiles to demonstrate the behavior shown by arrows in (d): When both the width and height of the channel are equal to 5 (a), the emergent flow is coherent and pumps fluid across the channel. When the width is increased to 20 (b), the flow loses coherence, and the average flow across the channel, $\braket{u_x}$ drops to zero. When the height is also increased to 20 (c), coherence is restored. \textbf{(d,e)} Heatmaps of average flow across the channel as a function of its dimensions for activity magnitude $\alpha=5$ (d) and $\alpha=7$ (e). The solid white line corresponds to the transition from stationary to flowing states predicted by the linear stability analysis. The dashed white lines encompass the region of coherent flows predicted by vortex sizes computed in (Fig. \ref{corrlLenfig}d).}\label{reentrantfig}
\end{figure}

\emph{Self-pumping in 3D.} To study the emergence of self-pumping flows in 3D, we assume a simple channel geometry with no slip walls at $y=0,W$ and $z=0,H$ (Fig. \ref{schematic}c), and no preferential anchoring (Neumann boundary conditions on $\bm{Q}$). The minimum channel dimensions that give rise to spontaneous flows are predicted by the stability analysis (solid white lines in Fig.\ref{reentrantfig}d,e). In channels with square cross-sections, strong confinements above the instability yield self-pumping flows that are uniform along the length of the channel (Fig.~\ref{reentrantfig}a). The nematic order is high near the walls and low in the middle of the channel, while the nematic director forms an oblique angle with the walls (SI), in agreement with experiments \cite{Wu2017}. An alternate pathway to this uniform self-pumping state is through the destabilization of a weakly ordered nematic state by a combination of fluctuations in the degree of nematic order and splay deformations, as shown by the stability analysis in \cite{Chandrakar2020}. The phenomenology observed on increasing both the channel dimensions commensurately is analogous to the behavior of a 2D confined system--- the flows develop components perpendicular to the channel axis at weaker confinements (Fig.~\ref{reentrantfig}c), and gradually lose their self-pumping nature (diagonals of Fig.~\ref{reentrantfig}d,e).
 
Increasing the channel dimensions \emph{incommensurately} reveals an intriguing behavior that is unique to 3D. Starting from a strongly confined system with a symmetric cross-section and uniform coherent flow (Fig.~\ref{reentrantfig}a), increasing either channel dimension destroys the self-pumping nature of the flow (Fig.\ref{reentrantfig}b). An analogy to 2D systems would suggest that the emergent flows continue to lose coherence as the channel dimensions are increased \cite{Chandragiri2020}. However, starting from a channel with a high aspect ratio cross-section (eg. Fig.\ref{reentrantfig}b), increasing the size of one dimension to lower the aspect ratio \emph{restores} self-pumping (Fig.\ref{reentrantfig}c). This is because weakening the confinement allows the flows to satisfy incompressibility by spilling into the third (z) dimension rather than by forming closed loops in the (xy) plane (SI). Note that the flows in this case are not uniform along the channel axis, and resemble the flows observed in experiments \cite{Wu2017}.
\begin{figure}
\centering
 \includegraphics[scale=0.14]{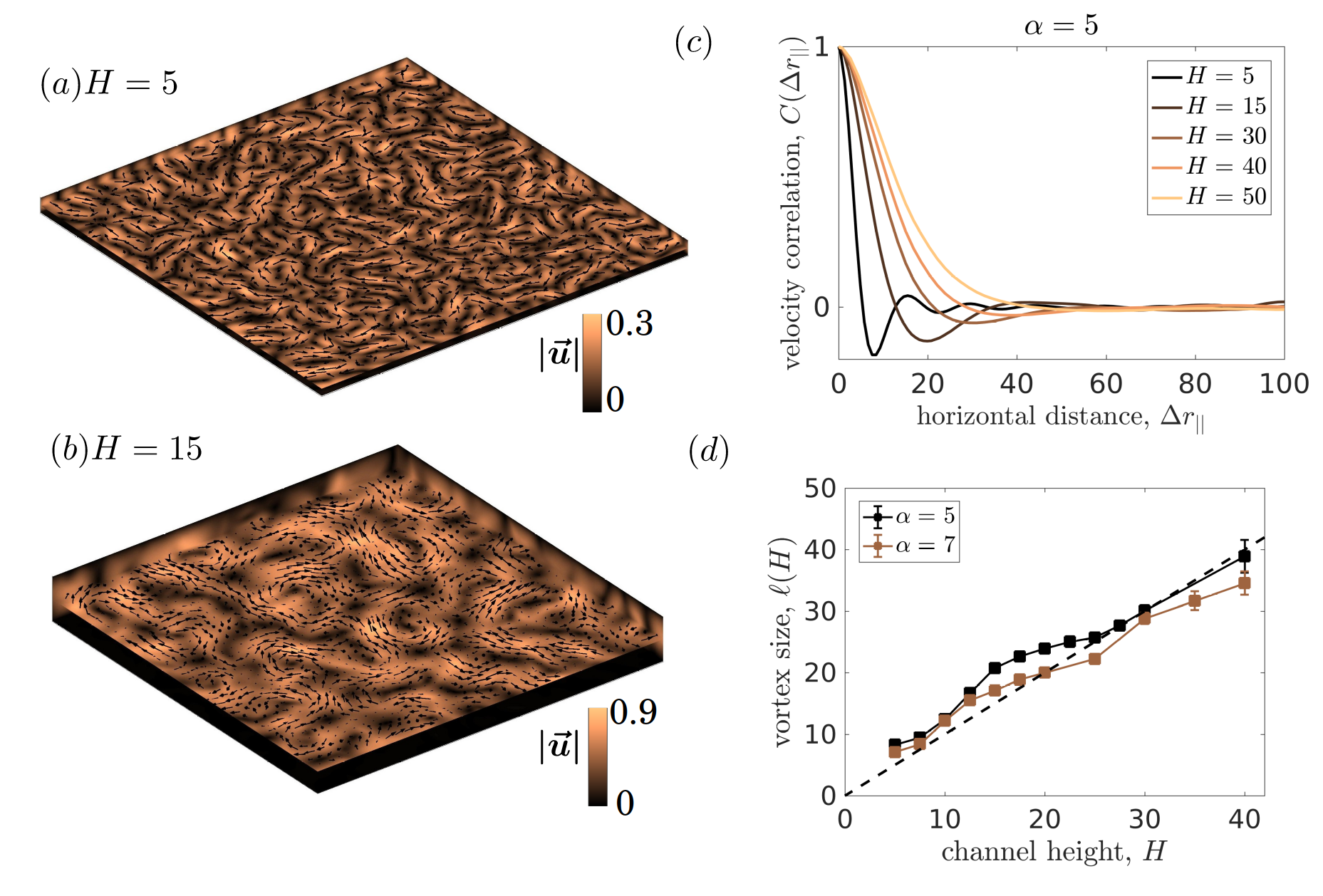}
 \caption{\textbf{(a, b)} Flow profile of an active fluid confined between two walls, showing that the vortex size increases with the separation between the walls, $H$. \textbf{(c)} The average velocity correlations as a function of distance in the horizontal plane, showing that the minimum of the correlation function shifts to the right (i.e., vortex size increases), as the distance between the walls is increased. At large values of $H$ (eg. H=50), there is no well defined vortex size. \textbf{(d)} Vortex size as a function of channel height. At small channel heights, the vortex size is larger than the channel height, but as the height increases, the vortex size falls below the height. }\label{corrlLenfig}
\end{figure}

To understand this novel effect of confinement in a 3D system, we consider a simpler confinement geometry: a pair of walls separated by a distance $H$ (Fig.~\ref{schematic}d). The emergent flows in this case are composed of swirling vortices, with a vortex size that increases with confinement size $H$ (Fig. \ref{corrlLenfig}a,b). A similar increase of vortex size with the confinement lengthscale has been reported in suspensions of microscopic swimmers \cite{Hernandez-Ortiz2009}, and in instabilities of nematically ordered systems \cite{Chandrakar2020}.

Our results show that coherent flows in channels emerge when the vortices generated by one pair of opposing walls are curtailed by the other pair of walls. Comparing the vortex sizes in Fig. \ref{corrlLenfig}d with the boundaries of the phase diagram for coherent flow in Fig. \ref{reentrantfig}d,e supports this argument: for all confinement geometries with coherent flows, the confinement length in each dimension is smaller than the size of the vortex induced by confinement in the other dimension. Since the vortex size increases with confinement length, this condition simplifies to the following condition for coherent flow: \emph{the larger confinement dimension has to be smaller than the size of the vortex induced by the smaller confinement dimension}. For example, consider a channel with width $W$ and height $H$, such that $W\geq H$. For one pair of channel walls separated by $H$ in the z direction, the vortex size in the xy plane is $\ell(H)$. We now introduce a new set of walls separated by $W$ in the y direction. If $W>\ell(H)$, at least one vortex can form in the xy plane, so the emergent flows will not be net pumping. On the other hand, if $W<\ell(H)$, there is not enough space to form a full vortex, so the flows are net pumping. Since $W\geq H$, coherent flow requires $\ell(H)>H$. Note that it is possible to have $\ell(H)<H$ (Fig. \ref{corrlLenfig}d) at large $H$. Therefore, there exists a maximum confinement length scale beyond which coherent flows cannot be obtained. For small activities (eg. $\alpha=5$), the confinement height at which $\ell(H)=H$ is nearly equal to the height at which there ceases to be a well-defined vortex size (Fig. \ref{corrlLenfig}c,d). For larger activities (eg. $\alpha=7$), $\ell(H)=H$ at a confinement height smaller than the height at which the minimum of the velocity correlation function disappears (Fig. \ref{corrlLenfig}d).

Experiments on bacterial suspensions confined in 2D channels have reported a similar requirement for existence of spontaneous flows; coherent flows were obtained when the width of the channel was smaller than the intrinsic bulk vortex size, while full vortices and no coherent flows resulted when the width of the channel was larger than this emergent length scale \cite{Wioland2016}. Note however, that even 3D channels with square cross-sections fundamentally differ from 2D channels: to obtain coherent flows, it is not sufficient that each confinement dimension is smaller than a bulk length scale. Rather, each confinement dimension has to be smaller than the corresponding confinement-induced length scale.

The negative velocity correlations and well-defined vortex sizes in our numerical results (Fig. \ref{corrlLenfig}c) arise from a complex nonlinear coupling between active flows and hydrodynamic screening. However, we can understand their system-size dependence by considering the following simple model. We approximate the nonlinear effects of flow on the dynamics of the nematic order as a stochastic Gaussian white noise $\Gamma_{\alpha\beta}$, so that $\partial_tQ_{\alpha\beta}=-\gamma Q_{\alpha\beta}+\kappa \nabla^2 Q_{\alpha\beta}+\sqrt{\Delta_\Gamma}\ \Gamma_{\alpha\beta}$. Let us compute the velocity correlations in the isotropic system confined within walls separated by $H$. Let $\vec{r}_{||}$ denote position in the xy plane, and $z$ denote position along the confinement dimension. Then, the velocity correlations in the unconfined directions are of the form (SI)
\begin{align}
&\lim_{|\vec{r}_{||}-\vec{r}'_{||}|\to\infty}\braket{u_\alpha(\vec{r}_{||},z,t)u_\alpha(\vec{r}'_{||},z,t)}_0\nonumber\\
&=\frac{\sqrt{2}\pi^2\alpha^2\Delta_\Gamma}{\gamma \eta^2} \sqrt{\frac{H}{|\vec{r}_{||}-\vec{r}'_{||}|}}\ \exp \left(-\frac{\pi|\vec{r}_{||}-\vec{r}'_{||}|}{H}\right) .
\end{align}
Thus, the velocity correlation length in the isotropic state is $H/\pi$. Above the flow alignment instability, the nonlinear dynamics give rise to a nontrivial dependence on confinement not captured by this simple analysis.

\emph{Conclusions.} In summary, we have shown that an isotropic suspension of extensile rod-like units develops spontaneous flows due to a hydrodynamic instability that couples extensile activity and shear alignment. The size of the vortices generated by this instability depends on the strength of confinement, and coherent flows arise when each confinement dimension is smaller than the size of vortices induced by confinement in the other dimension. This requirement results in aspect-ratio-dependent self-pumping, as observed in recent experiments \cite{Wu2017}. Crucially, the minimal nature of our model establishes that this phenomenon is generic to all shear-aligning extensile 3D active systems.

The range of channel sizes that allow net pumping flows can be tuned by controlling material parameters such as activity and diffusion rates, to regulate the velocity correlations. However, the nature of the flow depends on a competition between spontaneous flows generated by the flow alignment instability and destabilization of these flows due to the generic hydrodynamic instability \cite{Simha2002,Chandrakar2020}, and thus arise only in a `sweet spot' of material parameters. If activity or system size is too small, the system is below the flow alignment instability and there are no flows; if the activity or system size is too large, the flows are turbulent. If the material parameters are such that the correlation length $\ell(H)$ is smaller than the confinement size $H$ for all activities high enough to generate spontaneous flows, emergent flows can never be coherent.

Our numerical solutions suggest that the aspect-ratio-dependence of coherent flows can be roughly predicted by examining the effect of confinement in each dimension independently. This simple deconstruction is surprising, since active systems are typically highly sensitive to boundaries. Moreover, while the predicted aspect-ratio-dependence of coherent flow is consistent with existing experiments \cite{Wu2017}, experiments have not yet observed an upper limit to the overall size of the channel cross-section that allows coherent flow. Since this maximum size arises due to the predicted sublinear dependence of vortex size on the confinement lengthscale $\ell(H)$ (Fig.~\ref{corrlLenfig}d), experimentally measuring the size of the largest vortices as a function of confinement dimensions would be the next step toward testing the hydrodynamic theory. To maximize the generality of our conclusions, the hydrodynamic theory we presented here contains only the minimal ingredients for generating the behavior observed in experiments. The theory that most accurately models the experimental system may have additional features. For example, active forces resulting from higher-order gradients of the nematic order could lead to additional instabilities in highly confined geometries \cite{Maitra2018}, and contribute to the loss of coherence observed in high-aspect-ratio channels.

We note that while we were in the final stages of completing this manuscript, a related insightful but complementary paper appeared that also describes numerical computation of flow states in 3D channels \cite{Chandragiri2020}. There, the authors report a loss of coherence of flows when the width of the channel is increased at fixed height, yielding a phenomenology analogous to that of confined 2D active fluids. Our work provides a more comprehensive understanding of the dependence of coherent flow on channel aspect ratio, and thus shows that confined 3D active fluids are fundamentally different from confined 2D active fluids. Further, the analytical investigations described here elucidate the interplay between the system-size dependence of the correlation length and coherent flow. Finally, our work shows that the aspect ratio dependence of coherent flows depends on material parameters, and thus is not an immutable constraint.

\begin{acknowledgments}
We acknowledge support from the Brandeis Center for Bioinspired Soft Materials, an NSF MRSEC, DMR-1420382 and DMR-2011846 (MV, AB, MFH), DMR-1855914 (MFH), and BSF-2014279 (MV and AB). Computational resources were provided by NSF XSEDE allocation TG-MCB090163 (Stampede) and the Brandeis HPCC which is partially supported by DMR-MRSEC 2011486.
\end{acknowledgments}

\bibliography{AspectRatio.bib}
\bibliographystyle{apsrev4-1}

\onecolumngrid
\appendix
\section{Hydrodynamic model}\label{fullmodel}
\subsection{Derivation from microscopic dynamics}
Consider $N$ non-interacting ellipsoids, each of length $l$ and diameter $b$ suspended in a fluid in $d$ dimensions. The center of mass of ellipsoid $i$, $\vec{r}_i$, evolves as $\partial_t\vec{r}_i=\vec{u}(\vec{r}_i)+\sqrt{2\kappa}\ \vec{\eta}^T(\vec{r}_i)$, where $\vec{u}(\vec{r})$ is the fluid velocity at position $\vec{r}$, $\eta^T$ is a stochastic Gaussian white noise, and $\kappa$ is the translational diffusion constant. Let the axis of orientation of this ellipsoid, $\pm \hat{\nu}_i$ be defined along its axis of symmetry (length). If the flows have small Reynolds number, the time evolution of $\hat{\nu}_i$ is given by \cite{Taylor1923,Bretherton1962}
\begin{align}
\partial _{t}\hat{\nu}_i=\boldsymbol{\Omega }(\vec{r}_i)\cdot \hat{\nu}_i+\lambda \left( \bm{I}-\hat{\nu}_i\hat{\nu}_i\right) \cdot
\mathbf{E}(\vec{r}_i)\cdot \hat{\nu}_i\nonumber\\
\hspace{20pt}+\sqrt{\gamma/2} \left( \bm{I}-\hat{\nu}_i\hat{\nu}_i\right)\cdot \vec{\eta}^R(\vec{r}_i)\label{microdyn}
\end{align}
where $\Omega_{\alpha\beta}=\frac{1}{2}(\partial_\beta u_\alpha-\partial_\alpha u_\beta)$  and  $E_{\alpha\beta}=\frac{1}{2}(\partial_\beta u_\alpha+\partial_\alpha u_\beta)$ represent spatial variations in the flow, $\lambda=\frac{l^2-b^2}{l^2+b^2}$, $\vec{\eta}^R$ is a Gaussian white noise, and $\gamma$ is the rotational diffusion constant. 
Let $c(\vec{r},\hat{\nu},t)=\sum_{i=1}^N\delta (\vec{r}-\vec{r}_i(t))\delta(\hat{\nu}-\hat{\nu_i}(t))/N$ denote the probability that there is an ellipsoid at position  $\vec{r}$ oriented along $ \hat{\nu}$ at time $t$. Using standard coarse-graining \cite{Doi1988,VanKampen2007,Baskaran2010},
\begin{align}
\partial _{t}c\left(\vec{r},\hat{\nu},t\right) =-\vec{u}\cdot \vec{\nabla} c+\kappa \nabla^2 c-\left(\hat{\nu}
\times \frac{\partial }{\partial \hat{\nu}}\right) \cdot\left( \hat{\nu}\times \partial
_{t}\hat{\nu}\right) c +\frac{\gamma}{4}\left(\hat{\nu}%
\times \frac{\partial} {\partial \hat{\nu}}\right)^2 c\label{cdyn} .
\end{align}
The zeroeth order orientational moment, $\int d\hat{\nu}\ c(\vec{r},\hat{\nu},t)\equiv\rho(\vec{r},t)$. From eqn. \ref{cdyn}, $\partial_t\rho=-\vec{u}.\vec{\nabla}\rho+\kappa\nabla^2\rho$, so that $\rho$ remains uniform if it is initially uniform. We will therefore assume $\rho(\vec{r},t)\equiv \rho$ for our model. Since the ellipsoids have head-tail symmetry, $c(\vec{r},\hat{\nu},t)=c(\vec{r},-\hat{\nu},t)$. Therefore the first order orientational moment,  $\braket{\nu_\alpha(\vec{r})}=\int c(\vec{r},\hat{\nu},t) \nu_\alpha\ d\hat{\nu}=0$. The second moment of the orientational distribution function is related to the  nematic order in the system
\begin{align}
 \braket{\nu_\alpha \nu_\beta-\frac{1}{d}\delta_{\alpha\beta}}(\vec{r},t)&=\int d\hat{\nu}\ c (\vec{r},\hat{\nu},t) \left(\nu_{\alpha }\nu_{\beta }-\frac{1}{d}\delta_{\alpha\beta}\right)\nonumber\\
&\equiv \rho S(\vec{r},t)\ \left(n_\alpha(\vec{r},t) n_\beta(\vec{r},t)-\frac{1}{d}\delta_{\alpha\beta}\right)\equiv  \rho Q_{\alpha\beta}(\vec{r},t)
\end{align}
where $\bm{Q}$ is the order parameter that characterizes a nematically ordered state, $S$ is a measure of the magnitude of nematic ordering, and $\hat{n}$ is a unit vector (Note that $\bm{Q}$ vanishes in the isotropic state, i.e., when $c(\vec{r},\hat{\nu})=c(\vec{r},t)$). Then,
\begin{align}
 \rho\partial_t Q_{\alpha\beta}(\vec{r},t)=\int d\hat{\nu}\left( \nu_{\alpha }\nu_{\beta }-\frac{1}{d}\delta_{\alpha\beta}\right) \partial _{t}c\left( \vec{r},\hat{\nu},t\right)\label{Qfromc}
\end{align}

From equations \ref{cdyn} and \ref{Qfromc},
\begin{align}
 \rho\partial_t Q_{\alpha\beta}
&=- \rho \vec{u}\cdot \vec{\nabla}Q_{\alpha\beta}+\int d\hat{\nu}(\nu_\beta \partial_t\nu_\alpha c+\nu_\alpha \partial_t\nu_\beta c)\label{Qfromc1}
\end{align}

From equations \ref{microdyn} and \ref{Qfromc1},
\begin{align}
\partial_t Q_{\alpha\beta}&=-\vec{u}\cdot \vec{\nabla}Q_{\alpha\beta}+ \Omega_{\alpha\gamma}\braket{\nu_\gamma \nu_\beta}+\braket{\nu_\alpha \nu_\gamma}\Omega_{\beta\gamma}\nonumber\\
&\hspace{20pt}+\lambda(E_{\alpha\sigma} \braket{\nu_\sigma\nu_\beta}+\braket{\nu_\alpha\nu_\sigma} E_{\beta\sigma}-2\braket{\nu_\alpha\nu_\beta \nu_\gamma\nu_\sigma}E_{\gamma\sigma})
\end{align}
Therefore,
\begin{align}
&\partial_t Q_{\alpha\beta}-\vec{u}\cdot \vec{\nabla}Q_{\alpha\beta}+Q_{\alpha\gamma}\Omega_{\gamma\beta}- \Omega_{\alpha\gamma}Q_{\gamma\beta}\nonumber\\
&\hspace{20pt}=-\gamma Q_{\alpha \beta}+\kappa \nabla^2 Q_{\alpha \beta}+\frac{2\lambda}{d}E_{\alpha\beta}+\lambda(Q_{\alpha\gamma} E_{\gamma\beta}+E_{\alpha\gamma} Q_{\gamma\beta})-\frac{2\lambda}{\rho}\braket{\nu_\alpha\nu_\beta \nu_\gamma\nu_\sigma}E_{\gamma\sigma})\label{incl4th}    
\end{align}
To close the equations self-consistently
\begin{align}
\partial_t\bm{Q}+\vec{u}\cdot\vec{\nabla}\bm{Q}+\bm{Q}\cdot\bm{\Omega}-\bm{\Omega}\cdot\bm{Q}&=-\gamma \bm{Q} +\kappa \nabla^2\bm{Q}\nonumber\\
&\hspace{-40pt}+\lambda\left[\frac{2}{d}\bm{E}+\bm{Q}\cdot\bm{E}+\bm{E}\cdot\bm{Q}-\frac{2}{d}\ \text{tr}(\bm{Q}\cdot\bm{E})\bm{I}\right]\label{Qdynamics}
\end{align}
\subsection{Non-dimensionalization}
For numerical solutions of the hydrodynamics, we use a non-dimensional representation of eqn. (\ref{Qdynamics}), of the form

\begin{align}
&\partial_{t*}\bm{Q}+\vec{u}^*\cdot\vec{\nabla}_*\bm{Q}+\bm{Q}\cdot\bm{\Omega}^*-\bm{\Omega}^*\cdot\bm{Q}=-\gamma^* \bm{Q} +\kappa^* \nabla_*^2\bm{Q}\nonumber\\
&\hspace{15pt}+\lambda\left[\frac{2}{d}\bm{E}^*+\bm{Q}\cdot\bm{E}^*+\bm{E}^*\cdot\bm{Q}-\frac{2}{d}\ \text{tr}(\bm{Q}\cdot\bm{E}^*)\bm{I}\right] 
\end{align}
where the variables and operators with a `$^*$' have been non-dimensionalized. In particular, $\gamma^*=t_0 \gamma$, and $\kappa^*=t_0l_0^2$. $\gamma^*$ and $\kappa^*$ can be set to unity independently, giving a characteristic timescale and length scale for non-dimensionalization --- $t_0=1/\gamma$, and $l_0=\sqrt{\kappa/\gamma}$.

Further, non-dimensionalizing the driven Stoke's equation leads to
\begin{align}
 \eta^* \nabla_*^2\vec{u}^*-\vec{\nabla}_*p^*-\alpha^*\ \vec{\nabla}_*\cdot\bm{Q}&=0
\end{align}
where $\eta^*=\frac{\eta t_0l_0^{d-2}}{m_0}$ and  $\alpha^*=\frac{\alpha t_0}{\eta}$. Setting $\eta^*=1$ gives a unit of mass:
\begin{align}
 m_0=\eta t_0 l_0^{d-2}
\end{align}
Therefore, the only material parameters that remain, and that can be changed independently, are the non-dimensional activity $\alpha^*$, and the flow alignment parameter $\lambda$  

\section{Numerical scheme and parameter values}
To find steady states to better accuracy, we will incorporate few higher order terms into eqn. \ref{Qdynamics}. Our numerical method \cite{Varghese2021} starts from an isotropic state and solves eqn. (\ref{fullQdynamics}) using a semi-implicit finite difference time stepping scheme\cite{Zhao2016}. To solve the Stokes equation while ensuring incompressibility, we use a box smoothing algorithm on a staggered grid\cite{Vanka1986}.

Above the flow alignment instability, higher order terms are needed to stabilize the dynamics. Therefore, we modify eqn. \ref{Qdynamics} to include higher order terms in $\bm{Q}$,
\begin{small}
\begin{align}
&\partial_t\bm{Q}+\vec{u}\cdot\vec{\nabla}\bm{Q}+\bm{Q}\cdot \bm{\Omega}-\bm{\Omega}\cdot\bm{Q}
=-\gamma\Big[A\ \bm{Q}+B\left(\bm{Q}^2-\frac{1}{d}tr\bm{Q}^2\bm{I}\right)+C\ (\text{tr}\bm{Q}^2)\bm{Q}\Big]-\kappa\nabla^2\bm{Q}\nonumber\\
&\hspace{150pt}+\lambda\left[\frac{2}{d}\bm{E}+\bm{Q}\cdot\bm{E}+\bm{E}\cdot\bm{Q}-\frac{2}{d}\ \text{tr}(\bm{Q}\cdot\bm{E})\bm{I}\right]\label{fullQdynamics}
\end{align}
\end{small}
In 3D, we assume $A=\left(1-\frac{\rho}{3}\right)$, $B=-\rho$, and $C=\rho$, where $\rho$ is the number density of the rods, to be consistent with lyotropic liquid crystal literature. For $\rho<2.67$, $\bm{Q}=0$ is the only minimum of the free energy. For $2.67<\rho<2.7$, the isotropic state remains the global minimum, and the nematic state is metastable. For $2.7<\rho<3$, the nematic state is the global minimum, while the isotropic state is metastable. For $\rho>3$, the nematic state is the only free energy minimum. We choose $\rho=2.65$, which guarantees that there is only one stationary steady state, and that it is isotropic. In 2D, we assume $A=1-\rho$, and $C=(1+\rho)/\rho^2$ (the term that depends on $B$ vanishes), with $\rho=0.8$. We will assume that the rod-like units have negligible diameters, so that the shear alignment constant $\lambda=1$.
All the results presented have been expressed in units of $l_0=\sqrt{\kappa/\gamma A}$, $t_0=1/\gamma A$, and $m_0=\eta t_0l_0^{d-2}$.

\section{Linear Stability Analysis}
The only homogeneous state admitted by the model is $\bm{Q}=0,\ \vec{u}=0$. Consider fluctuations about this stationary isotropic state, $Q_{\alpha\beta}(\vec{r})=\int d\vec{k}\ \tilde{Q}_{\alpha\beta}(\vec{k})e^{-i\vec{k}\cdot\vec{r}}$, and $ u_{\alpha}(\vec{r})=\int d\vec{k}\  \tilde{u}_{\alpha}(\vec{k})e^{-i\vec{k}\cdot\vec{r}}$. Up to linear order in the fluctuation amplitudes, equation (\ref{Qdynamics}) gives
\begin{align}
\partial_t\tilde{Q}_{\alpha\beta}=-\gamma(A+\kappa k^2)\tilde{Q}_{\alpha\beta}-\frac{i\lambda}{d}(k_\alpha \tilde{u}_\beta+k_\beta \tilde{u}_\alpha)\label{Qlin}
\end{align}
Eliminating the velocity using
\begin{align}
 \tilde{u}_\beta=\frac{i\alpha \hat{k}_\nu}{\eta k}(\delta_{\beta \mu}-\hat{k}_\beta \hat{k}_\mu)\tilde{Q}_{\mu \nu}\label{uf}
\end{align}
gives
\begin{align}
 &\partial_t\tilde{Q}_{\rho\sigma}=-\gamma(A+\kappa k^2)\tilde{Q}_{\rho\sigma}\nonumber\\
 &+\frac{\lambda\alpha}{d\eta}\Big[\hat{k}_\sigma(\delta_{\rho\mu}-\hat{k}_\rho\hat{k}_\mu)+\hat{k}_\rho(\delta_{\sigma\mu}-\hat{k}_\sigma\hat{k}_\mu)\Big]\hat{k}_\nu\tilde{Q}_{\mu\nu}\label{Qlsa}
\end{align}
and
\begin{align}
 \partial_t\tilde{u}_\beta&=\frac{i\alpha \hat{k}_\nu}{\eta k}(\delta_{\beta \mu}-\hat{k}_\beta \hat{k}_\mu)\partial_t\tilde{Q}_{\mu \nu}\nonumber\\
 &=\Big[-\gamma(A+\kappa k^2)+\frac{\lambda\alpha}{d\eta}\Big]\tilde{u}_\beta\label{LSA_u} .
\end{align}
The physics of these linearized dynamics is discussed in the main text.
\newpage
\section{Phenomenology in 2D}
\begin{figure}[H]
\centering
 \includegraphics[scale=0.25]{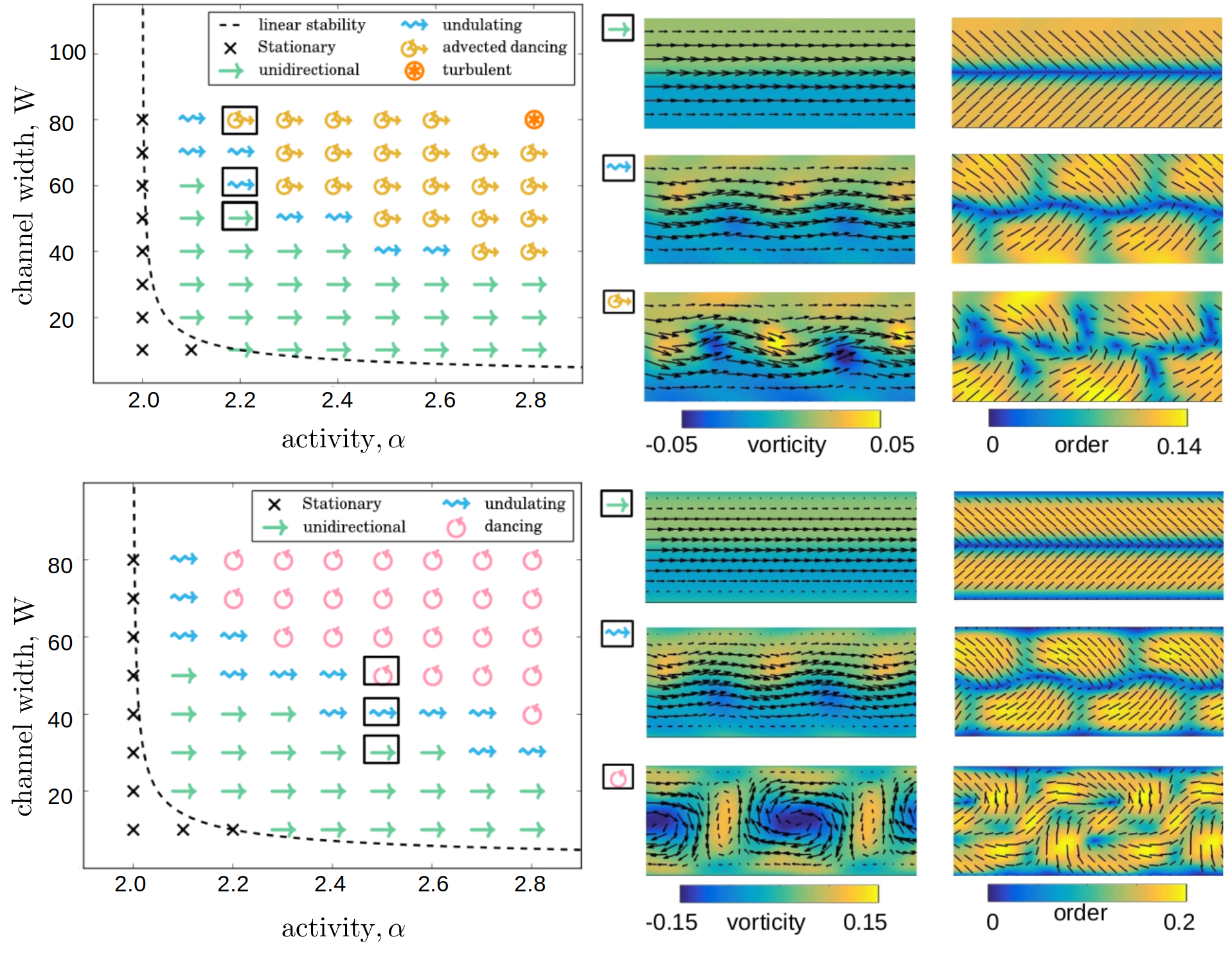}
 \caption{Flow states of a confined 2D incipient active nematic under (top panel) no anchoring and (bottom panel) parallel anchoring boundary conditions. In each case, the left panel gives the phase behavior as a function of activity and channel width. For the highlighted points in the phase diagram, the right panel shows the flow and nematic order profiles. In each case, crossing the phase boundary predicted by the flow alignment instability gives rise to unidirectional flow along the channel. This flow is produced by a splay wall-like structure in a weakly ordered background. As the channel width is increased, the bend instability gives rise to an undulation in the splay wall and a corresponding undulation in the flow. When the confinement is weakened further, the splay wall unzips into $\pm 1/2$ defects, with the $+1/2$ defects performing dancing dynamics as in \cite{Shendruk2017}. For free orientations at the boundaries, this dancing dynamics occurs in a background net-pumping flow. This arises because, when there is no anchoring, the flow gradient due to no slip builds a finite order at the walls, with the director aligned obliquely at the walls. On enforcing parallel anchoring at the walls, flow alignment sharply bends the director near the wall to form an oblique angle, so there is no order near the walls.}
\end{figure}
\newpage
\section{Phenomenology in 3D}
\begin{figure}[H]
    \centering
    \includegraphics[scale=0.2]{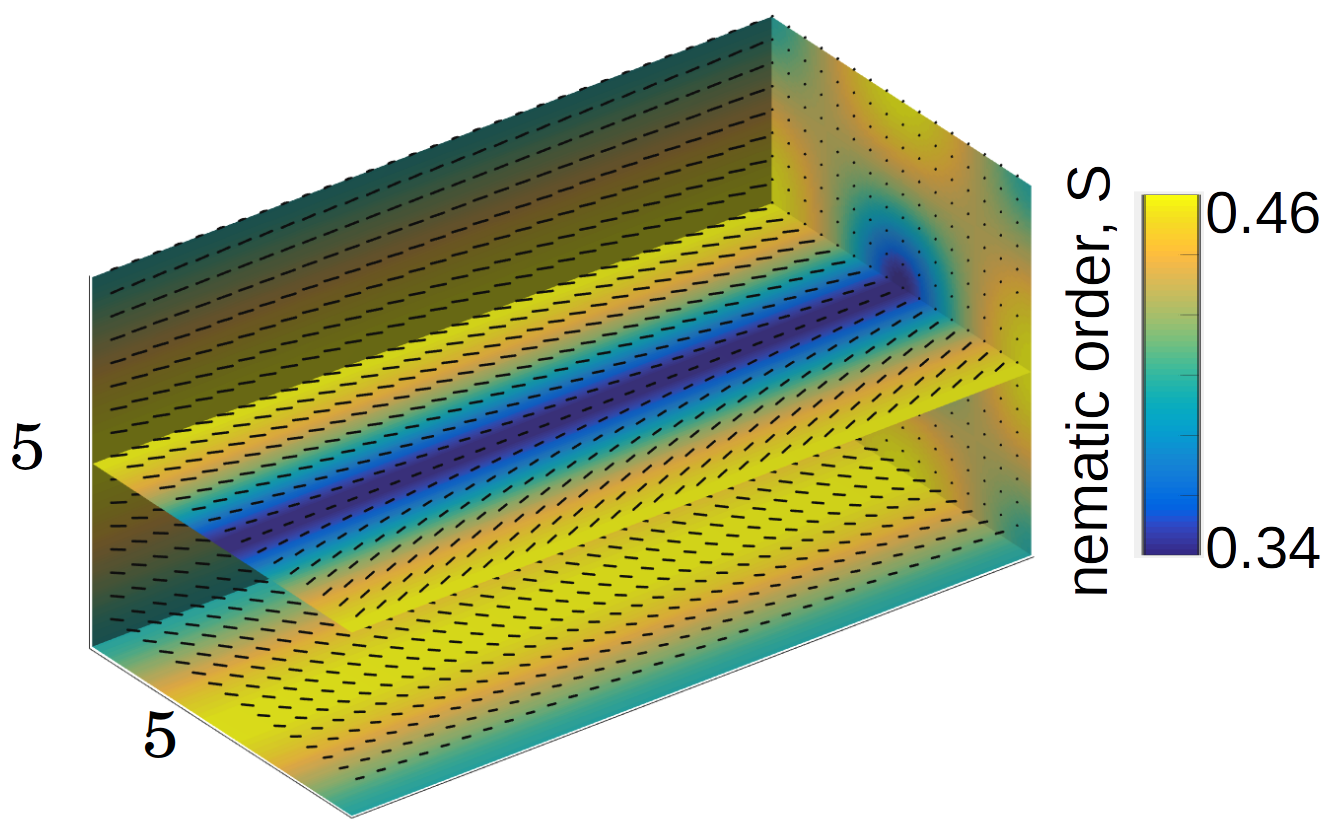}
    \caption{Degree of nematic order (colormap) and orientation field (black lines) that gives rise to uniform coherent flows in a strongly confined system. There is a high degree of nematic order near the walls, and a low degree of nematic order in the middle of the channel. The nematic orientation has a splay-like profile, with the orientation at the walls forming an oblique angle with the walls.}
    \label{splay_wall}
\end{figure}
\begin{figure}[H]
    \centering
    \includegraphics[scale=0.25]{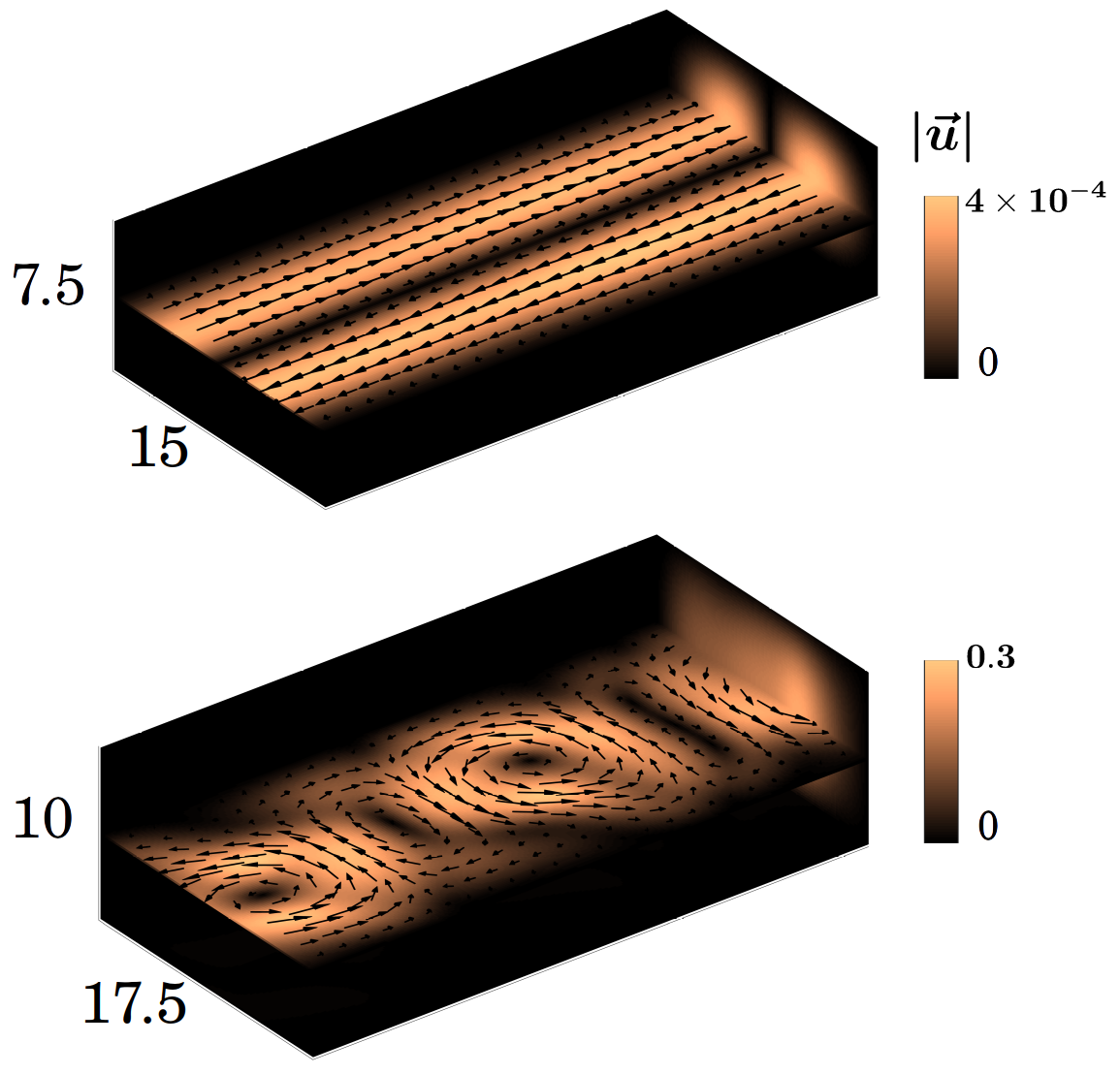}
    \caption{Flow profiles in asymmetric channels with planar anchoring. \textbf{(top panel)} In asymmetric channels that are strongly confined along one dimension, coherent flow is lost due to formation of banded flows. \textbf{(bottom panel)} In weakly confined asymmetric channels, the flows consist of vortex arrays, like in the case without anchoring at the walls. The phenomenology for homeotropic anchoring is identical to that for planar anchoring.}
    \label{anchoring}
\end{figure}
\begin{figure}[H]
    \centering
    \includegraphics[scale=0.2]{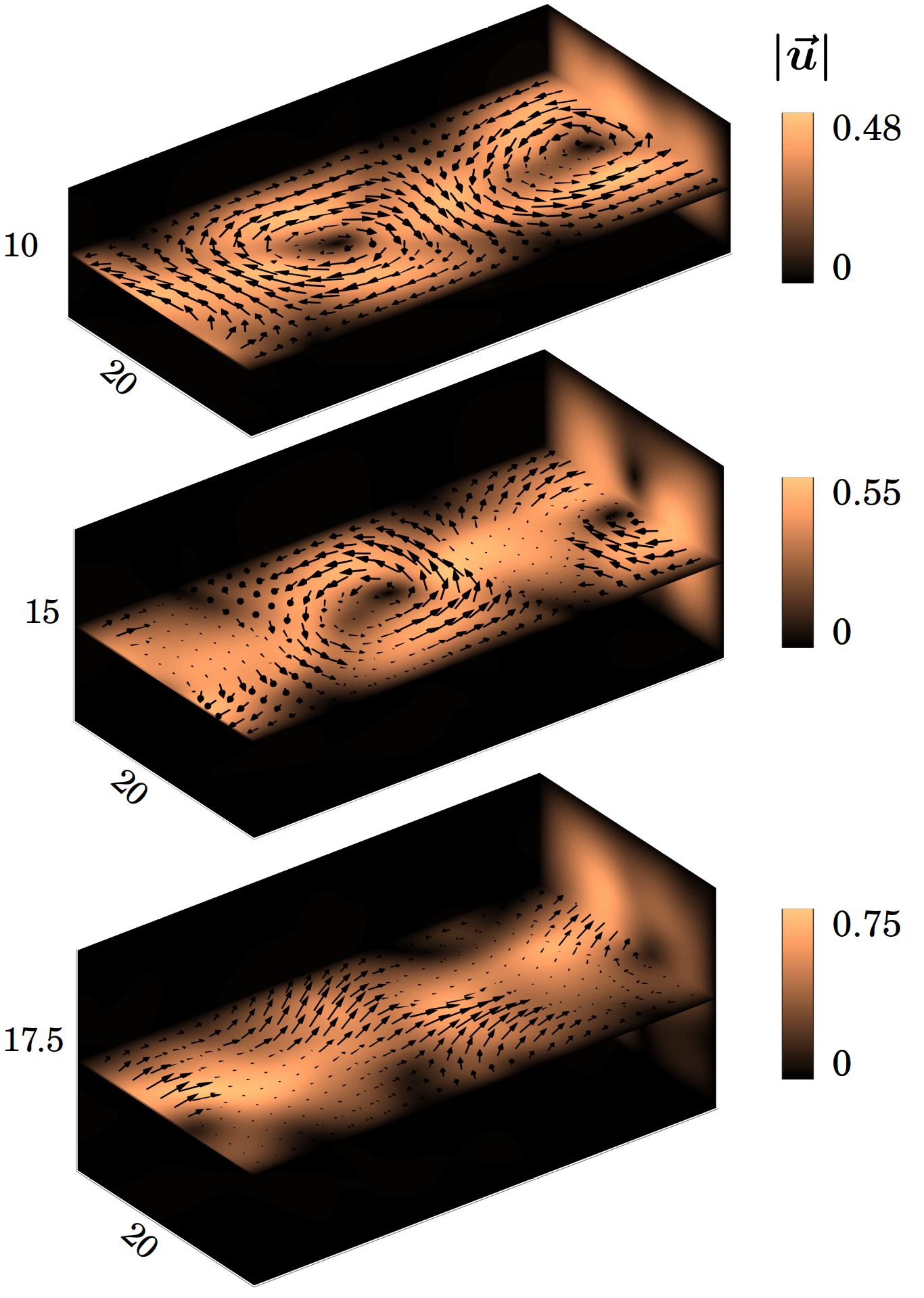}
    \caption{Increase in coherence of the emergent flows when the channel height is increased to match the width: increasing the height allows the flows to satisfy incompressibility by spilling into the third (z) dimension rather than by forming closed loops in the (xy) plane}
    \label{increaseH}
\end{figure}
\begin{figure}[H]
    \centering
    \includegraphics[scale=0.2]{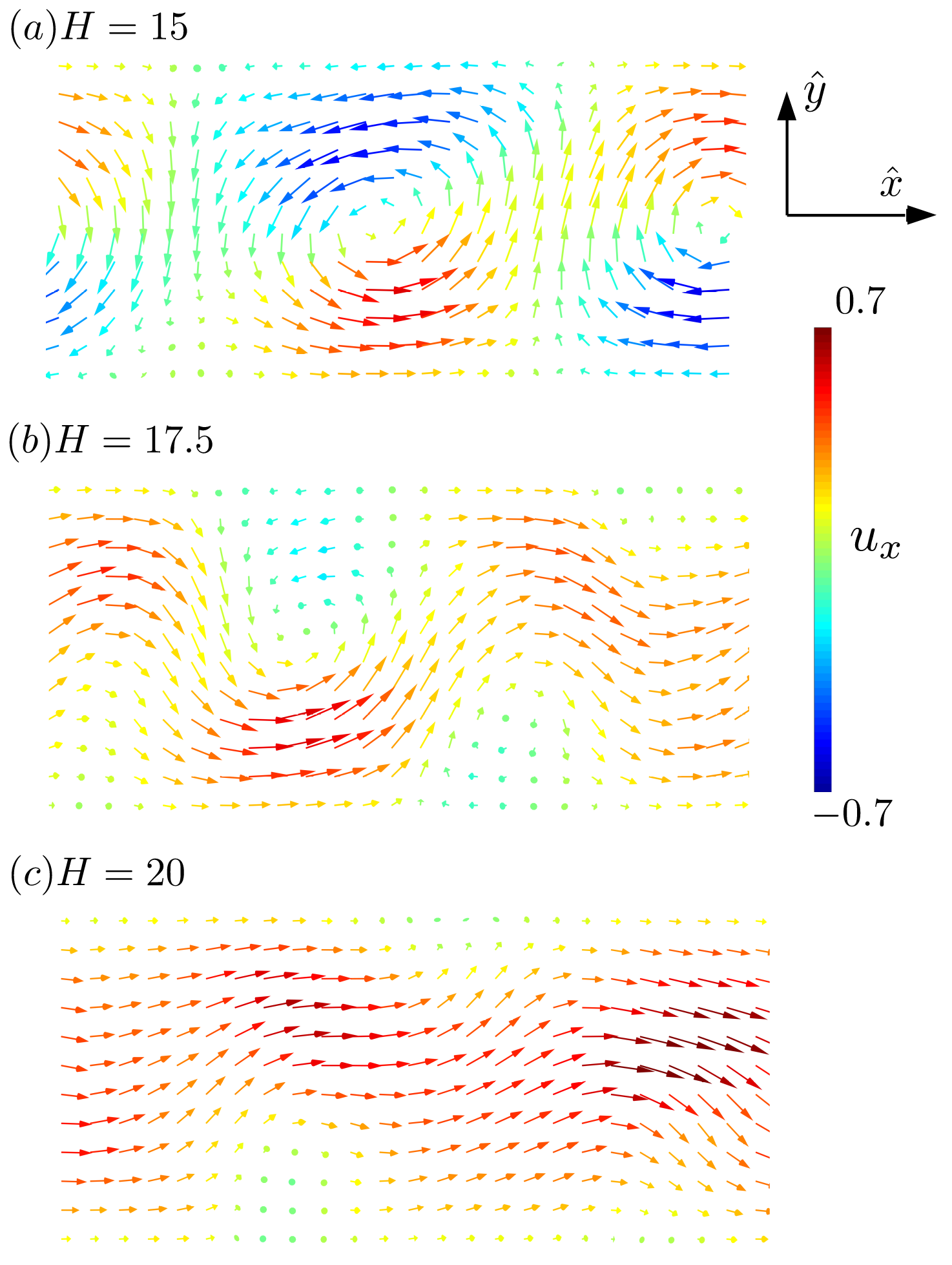}
    \caption{View of velocity profile in the XY plane at z=H/2 and fixed width W=20. The colors of the velocity arrows correspond to the velocity component along the channel axis ($\hat{x}$). For H=15, the flows consist of fully formed vortices in the XY plane. As the channel height is increased, flow along one of the directions gains prominence, giving rise to flows that pump the fluid along the channel (coherent flow).}
    \label{increaseH2}
\end{figure}
\section{Velocity Correlations}
\subsection{Definition}
We calculate the velocity correlations in the unconfined plane as
\begin{align}
 C(\Delta r_{||})=\frac{\int_0^H dz\int d\vec{r}_{||}\int d\vec{r}'_{||}\ \vec{u}(\vec{r}_{||},z)\cdot \vec{u}(\vec{r}_{||}',z)\delta(\vec{r}_{||}-\vec{r}'_{||})}{\int_0^H dz\int d\vec{r}_{||}\vec{u}(\vec{r}_{||},z)\cdot \vec{u}(\vec{r}_{||},z)}\label{Crpll}
\end{align}
The vortex size is the distance over which this planar velocity correlation function(Fig. 2(c)) becomes maximally anticorrelated, i.e., reaches its minimum value. Note that the vortex size calculated by averaging over all the horizontal planes, as in eqn.~\ref{Crpll}, is nearly equal to the vortex size calculated within the midplane ($z=H/2)$. This occurs because: although the behavior of the planar velocity correlations depend strongly on the distance of the measurement plane from the walls (vortex size increases with distance from the walls and attains its maximum value at $z=H/2$),  the most significant contribution to the $z$ integrals in eqn. (\ref{Crpll}) come from $z=H/2$. This is because at the strong confinements considered here, the velocity drops quickly away from the $z=H/2$ plane. Also, the $x$ and $y$ components of the velocity are much larger than the $z$ component at strong confinement. Therefore, the vortex sizes calculated by eqn. (\ref{Crpll}) are nearly equivalent to those computed from planar components of the velocity in the $z=H/2$ plane.
\subsection{Estimate of velocity correlations in the isotropic state}
 Suppose the nonlinear complex effects of flow on the dynamics of the nematic order can be approximated as a stochastic Gaussian white noise $\Gamma_{\alpha\beta}$. Then,
 \begin{equation}
 \partial_tQ_{\alpha\beta}=-\gamma Q_{\alpha\beta}+\kappa \nabla^2 Q_{\alpha\beta}+\sqrt{\Delta_\Gamma}\ \Gamma_{\alpha\beta} \label{dtQ}   
 \end{equation} 
 where
 \begin{small}
 \begin{align}
    \braket{ \tilde{\Gamma}_{\alpha\beta}(\vec{k}_{||},n,\omega)\tilde{\Gamma}_{\mu\nu}(\vec{k}'_{||},n',\omega')}&= \delta^2(\vec{k}_{||}+\vec{k}'_{||})\delta_{nn'}\delta(\omega+\omega')(\delta_{\alpha\mu}\delta_{\beta\nu}+\delta_{\alpha\nu}\delta_{\beta\mu}-\frac{2}{3}\delta_{\alpha\beta}\delta_{\mu\nu}) .
    \label{Gamma}
 \end{align}
 \end{small}
 Consider confinement within walls separated by a distance $H$. Let $\vec{r}_{||}$ denote position in the xy plane, and $z$ denote position along the confinement dimension. Then,
 \begin{align*}
 Q_{\alpha\beta}(\vec{r}_{||},z,t)=\int_{-\infty}^\infty d\omega e^{i\omega t}\int_{0}^\infty dk_{||}k_{||}\int_0^{2\pi} d\theta e^{-ik_{||}r_{||}\cos\theta}\sum_{n=1}^\infty e^{\frac{in\pi z}{H}}\ \tilde{Q}_{\alpha \beta}(\vec{k}_{||},n,t).
\end{align*}
 From equations \ref{dtQ} and \ref{Gamma}, 
 \begin{align}
 \braket{\tilde{Q}_{\alpha\beta}(\vec{k},\omega)\tilde{Q}_{\mu\nu}(-\vec{k},-\omega)}&=\frac{\braket{\tilde{\Gamma}_{\alpha\beta}(\vec{k},\omega)\tilde{\Gamma}_{\mu\nu}(-\vec{k},-\omega)}}{\omega^2+(\gamma+\kappa k^2)^2}\nonumber\\
 &=\frac{\Delta_\Gamma}{\omega^2+(\gamma+\kappa k^2)^2}[\delta_{\alpha\mu}\delta_{\beta\nu}+\delta_{\alpha\nu}\delta_{\beta\mu}-\frac{2}{3}\delta_{\alpha\beta}\delta_{\mu\nu}] .\nonumber
 \end{align}
 In particular,
 \begin{align}
      \braket{\tilde{Q}_{\alpha\beta}(\vec{k},\omega)\tilde{Q}_{\mu\nu}(-\vec{k},-\omega)}&=\frac{10\ \Delta_\Gamma}{\omega^2+(\gamma+\kappa k^2)^2},
 \end{align}
 giving
\begin{align}
 &\braket{Q_{\alpha\beta}(\vec{r}_{||},z,t)Q_{\alpha\beta}(\vec{r}'_{||},z,t)}\nonumber\\
 &=\sum_{n=1}^\infty e^{\frac{in\pi (0)}{H}}\int dk_{||}k_{||}\int_0^{2\pi} d\theta e^{ik_{||}\Delta r_{||}\cos\theta} \braket{\tilde{Q}_{\alpha\beta}(\vec{k})\tilde{Q}_{\alpha\beta}(-\vec{k})}\nonumber\\
  &=\sum_{n=1}^\infty\int dk_{||} J_0(k_{||}\Delta r_{||})\ k_{||}\braket{\tilde{Q}_{\alpha\beta}(\vec{k})\tilde{Q}_{\alpha\beta}(-\vec{k})}\nonumber\\
 &=10\sum_{n=1}^\infty \int_{0}^\infty dk_{||}k_{||} J_0(k_{||}\Delta r_{||}) \int_0^\infty d\omega\frac{\Delta_\Gamma}{\omega^2+(\gamma+\kappa k^2)^2}\nonumber\\
 &=10\pi^2\Delta_\Gamma\int_{0}^\infty dk_{||}k_{||}J_0(k_{||}\Delta r_{||})\sum_{n=1}^\infty \frac{1}{\gamma+\kappa k^2}\nonumber\\
 &=\frac{10\pi^2\Delta_\Gamma}{\kappa}\sum_{n=1}^\infty\int_{0}^\infty dk_{||}k_{||}J_0(k_{||}\Delta r_{||}) \frac{1}{k_{||}^2+\frac{\gamma}{\kappa}+  \frac{\pi^2 n^2}{H^2}}\nonumber\\
 &=\frac{10\pi^2\Delta_\Gamma}{\kappa}\sum_{n=1}^\infty K_0\left(\Delta r_{||}\sqrt{\frac{\gamma}{\kappa}+\frac{\pi^2 n^2}{H^2}}\right)
\end{align}
where $J_0(x)$ is the zeroeth Bessel function of the first kind, and $K_0(x)$ is the modified Bessel function of the second kind, which can be approximated as
\begin{align}
 \lim_{x\to\infty}K_0(x)=\sqrt{\frac{\pi}{2x}}e^{-x}\left(1+O\left(\frac{1}{x}\right)\right) .
\end{align}
Therefore,
\begin{align}
 \lim_{\Delta r_{||}\to \infty}\braket{Q_{\alpha\beta}(\vec{r}_{||},z,t)Q_{\alpha\beta}(\vec{r}'_{||},z,t)}=\sqrt{ \frac{\pi}{2\Delta r_{||}}}\left(\frac{\gamma}{\kappa}+\frac{\pi^2 n^2}{H^2}\right)^{-\frac{1}{4}}e^{-\Delta r_{||}\sqrt{\frac{\gamma}{\kappa}+\frac{\pi^2 n^2}{H^2}}},
\end{align}
and thus the nematic order correlation in the isotropic state, when confined between planes separated by a distance $H$ is, $l_Q=\sqrt{\frac{\kappa}{\gamma}}\frac{H}{\pi^2+H^2}$.

The velocity correlations in the isotropic state can then be calculated as
\begin{align}
 &\braket{\tilde{u}_\alpha(\vec{k},\omega)\tilde{u}_\alpha(-\vec{k},-\omega)}\nonumber\\
 &=+\frac{\alpha^2}{\eta^2 k^2}\hat{k}_\nu\hat{k}_\sigma\Big[\delta_{\alpha \mu}-\hat{k}_\alpha \hat{k}_\mu\Big]\Big[\delta_{\alpha \gamma}-\hat{k}_\alpha \hat{k}_\gamma\Big]\braket{\tilde{Q}_{\mu\nu}(\vec{k},\omega)\tilde{Q}_{\gamma\sigma}(-\vec{k},-\omega)}\nonumber\\
 &=\frac{\alpha^2\Delta_\Gamma}{\eta^2 k^2}\ \hat{k}_\nu\hat{k}_\sigma
 \Big[\delta_{\mu\gamma}-\hat{k}_\mu\hat{k}_\gamma\Big]\frac{[\delta_{\mu\gamma}\delta_{\nu\sigma}+\delta_{\mu\sigma}\delta_{\nu\gamma}-\frac{2}{3}\delta_{\mu\nu}\delta_{\gamma\sigma}]}{\omega^2+(\gamma+\kappa k^2)^2}\nonumber\\ 
 &=\frac{\alpha^2\Delta_\Gamma}{\eta^2 k^2}
\frac{(2+0-\frac{2}{3}\times0)}{\omega^2+(\gamma+\kappa k^2)^2}\nonumber\\ 
&=\frac{\alpha^2\Delta_\Gamma}{\eta^2 k^2}\ \frac{2}{\omega^2+(\gamma+\kappa k^2)^2}
\end{align}
\begin{align}
 &\braket{u_\alpha(\vec{r}_{||},z,t)u_\alpha(\vec{r}'_{||},z,t)}\nonumber\\
  &=\sum_{n=1}^\infty\int dk_{||} J_0(k_{||}\Delta r_{||})\ k_{||}\braket{\tilde{u}(\vec{k})\tilde{u}(-\vec{k})}\label{ucorrl_J1}\\
 &=2\pi\sum_{n=1}^\infty\int_{0}^\infty dk_{||}k_{||} J_0(k_{||}\Delta r_{||})\int_0^\infty d\omega \Big[\frac{\alpha^2\Delta_\Gamma}{\eta^2 k^2}\ \frac{2}{\omega^2+(\gamma+\kappa k^2)^2}\Big]\nonumber\\
 &=\frac{2\pi^2\alpha^2\Delta_\Gamma}{\kappa\eta^2}\int_{0}^\infty dk_{||}k_{||}J_0(k_{||}\Delta r_{||}) \sum_{n=1}^\infty\frac{1}{k_{||}^2+\frac{\pi^2n^2}{H^2}}\ \frac{1}{\frac{\gamma}{\kappa}+k_{||}^2+\frac{\pi^2n^2}{H^2}}\label{ucorrl_J2}\\
 &=\frac{2\pi^2\alpha^2\Delta_\Gamma}{\gamma\eta^2}\sum_{n=1}^\infty\left[K_0(\Delta r_{||}\frac{n\pi}{H})-K_0(\Delta r_{||}\sqrt{\frac{\gamma}{\kappa}+\frac{n^2\pi^2}{H^2}})\right]
 \end{align}
Therefore,
\begin{align}
\lim_{|\vec{r}_{||}-\vec{r}'_{||}|\to\infty}\braket{u_\alpha(\vec{r}_{||},z,t)u_\alpha(\vec{r}'_{||},z,t)}=\frac{\sqrt{2}\pi^2\alpha^2\Delta_\Gamma}{\gamma \eta^2} \sqrt{\frac{H}{|\vec{r}_{||}-\vec{r}'_{||}|}}\ \exp \left(-\frac{\pi|\vec{r}_{||}-\vec{r}'_{||}|}{H}\right)
\end{align}
Thus, the velocity correlation length in the isotropic state is $H/\pi$.

Consider the expression in eqn. \ref{ucorrl_J1} for the velocity correlation function. In order to obtain correlation functions of the kind in Fig. 3(c) of the main text, that preserve the oscillatory nature of $J_0(x)$ and give a well defined vortex size,   $k_{||}\braket{\tilde{u}(\vec{k})\tilde{u}(-\vec{k})}$ should be peaked around a certain value of $k_{||}$ and this peak should be sufficiently narrow. In the expression in eqn. \ref{ucorrl_J2},  
\begin{align}
 k_{||}\braket{\tilde{u}(\vec{k})\tilde{u}(-\vec{k})}&\sim \left(\frac{k_{||}}{\frac{1}{H^2} +k_{||}^2}\right)\frac{1}{\frac{\gamma}{\kappa}+\frac{1}{H^2} +k_{||}^2} ,
 \label{ucork}
\end{align}
the peak of this function is too broad to retain the oscillatory nature of $J_0(x)$. Therefore, the velocity correlations in the isotropic state do not have the oscillatory behavior seen above the flow aligning transition (main text, Fig. 3c). We therefore conclude that the emergent length scale observed in the numerical solutions arises from a characteristic length scale of correlations in the nematic order, and that this length scale arises from active nonlinear processes.
\end{document}